\documentstyle[11pt,aaspp4]{article}

\newcommand{\lii}{Li\,{\footnotesize I}}
\newcommand{\fei}{Fe\,{\footnotesize I}}
\newcommand{\kms}{\,km\,s$^{-1}$}
\newcommand{\vsini}{\mbox{$v_e\,\sin\,i$}}

\lefthead{Jeffries \& James}
\righthead{Lithium in Blanco 1}

\begin{document}

\title{Lithium in Blanco 1: Implications for Stellar Mixing}

\author{Robin D. Jeffries}
\affil{Department of Physics, Keele University, Keele, Staffordshire,
ST5 5BG, UK}

\and

\author{David J. James}
\affil{School of Physics and Astronomy, University of St. Andrews, North
Haugh, St. Andrews, Fife, KY16 9SS, UK}

% Notice that each of these authors has alternate affiliations, which
% are identified by the \altaffilmark after each name.  The actual alternate
% affiliation information is typeset in footnotes at the bottom of the
% first page, and the text itself is specified in \altaffiltext commands.
% There is a separate \altaffiltext for each alternate affiliation
% indicated above.

% The abstract environment prints out the receipt and acceptance dates
% if they are relevant for the journal style.  For the aasms style, they
% will print out as horizontal rules for the editorial staff to type
% on, so long as the author does not include \received and \accepted
% commands.  This should not be done, since \received and \accepted dates
% are not known to the author.

\begin{abstract}
We obtain lithium abundances for G and K stars in Blanco 1, an open
cluster with an age similar to, or slightly younger than, the Pleiades.
We critically examine previous spectroscopic abundance analyses of
Blanco 1 and conclude that while there were flaws in earlier work, it
is likely that Blanco 1 is close in overall metallicity to the older
Hyades cluster and more metal-rich than the Pleiades.  However, we find
Blanco 1 has Li abundances and rotation rates similar to the Pleiades,
contradicting predictions from standard stellar evolution models, in
which convective pre-main sequence (PMS) Li depletion should increase
rapidly with metallicity. If the high metallicity of Blanco 1 is
subsequently confirmed, our observations imply (1) that a currently
unknown mechanism severely inhibits PMS Li depletion, (2) that
additional non-standard mixing modes, such as those driven by rotation
and angular momentum loss, are then responsible for main sequence Li
depletion between the ages of Blanco 1 and the Hyades, and (3) that in
clusters younger than the Hyades, metallicity plays only a minor role
in determining the amount of Li depletion among G and K stars. These
conclusions suggest that Li abundance remains a useful age indicator
among young ($<700$\,Myr) stars even when metallicities are unknown. If
non-standard mixing is effective in Population\,I stars, the primordial
Li abundance could be significantly larger than present day
Population\,II Li abundances, due to prior Li depletion.

\end{abstract}

% The different journals have different requirements for keywords.  The
% keywords.apj file, found on aas.org in the pubs/aastex-misc directory, 
% contains a list of keywords used with the ApJ and Letters.  These are 
% usually assigned by the editor, but authors may include them in their 
% manuscripts if they wish. 


% That's it for the front matter.  On to the main body of the paper.
% We'll only put in tutorial remarks at the beginning of each section
% so you can see entire sections together.

% In the first two sections, you should notice the use of the LaTeX \cite
% command to identify citations.  The citations are tied to the
% reference list via symbolic KEYs.  We have chosen the first three
% characters of the first author's name plus the last two numeral of the
% year of publication.  The corresponding reference has a \bibitem
% command in the reference list below.
%
% Please see the AASTeX manual for a more complete discussion on how to make
% \cite-\bibitem work for you.   

\section{Introduction}

Lithium is of great importance in the study of stellar interiors, and
cosmology. The primordial Li abundance, Li$_{\rm p}$, constrains models
of big bang nucleosynthesis and the value of the baryonic density ({\em
e.g.}  Boesgaard \& Steigman 1985, Bonifacio \& Molaro 1997).  Li$_{\rm
p}$ is estimated to be $A({\rm Li}) = 12 + \log [N({\rm Li})/N({\rm
H})] = 2.2-2.3$, from observations of Li in Population\,II stars, {\em
if} the observed Li is almost unaltered from Li$_{\rm p}$ ({\em e.g.}
Spite \& Spite 1982, Bonifacio \& Molaro 1997).  $A({\rm Li})$ in the
youngest stars and in meteorites is 3.2-3.3 and while ongoing Galactic
Li enrichment processes are possible, there is growing evidence that Li
in Population\,II stars has been {\em depleted} from Li$_{\rm p}$
(Deliyannis \& Ryan 1997, Boesgaard et al. 1998).  Depletion could
occur because Li is destroyed in p,$\alpha$ reactions at only
$2.4\times10^{6}$\,K. ``Standard'' stellar evolution models,
incorporating only convective mixing, predict almost no Li depletion in
Population\,II stars ({\em e.g.} Deliyannis, Demarque \& Kawaler 1990),
because of their low metallicities and thin convection zones
(CZs). However, models incorporating ``non-standard'' mixing, such as
microscopic diffusion and turbulence driven by angular momentum loss
and differential rotation, predict significant Li depletion
(Pinsonneault, Deliyannis \& Demarque 1992, Chaboyer \& Demarque 1994).
Li$_{\rm p}$ could then be an order of magnitude greater than present
day Population\,II Li abundances.

Observing Li in the cool stars of young open clusters is an excellent
way of exploring the physics of non-standard stellar mixing. Li
enrichment processes can be neglected and rapid AML as stars reach the
ZAMS is likely to make Li depletion due to non-standard processes more
dramatic (Pinsonneault 1997 [P97]).  Standard models reproduce the
decline of Li with decreasing mass and increasing CZ depth in open
clusters, but have difficulty explaining the spread in Li abundance
among the cool stars of the Pleiades (age 100\,Myr -- Meynet,
Mermilliod, \& Maeder 1993), or why Li depletion apparently continues
between the Pleiades and Hyades (age 700\,Myr), despite CZ bases being
too cool to burn Li in G/K stars once they reach the ZAMS (Soderblom et
al. 1993 [S93], Thorburn et al. 1993 [T93]). One hypothesis is that
revised OPAL opacities (Rogers \& Iglesias 1992) and the metallicity
difference between the Pleiades and Hyades ([Fe/H]$=-0.034\pm0.024$ for
the Pleiades and $+0.127\pm 0.022$ for the Hyades -- Boesgaard \& Friel
1990 [BF90]) can explain the extra Li depletion and that no
non-standard main-sequence (MS) mixing is required (Swenson et
al. 1994a,b [S94a,b]). Increased metallicity leads to deeper CZs and
increased standard {\em pre}-main sequence (PMS) Li depletion in the
Hyades. Apparent scatter in the Pleiades Li abundances may be caused by
small metallicity differences or an imperfect treatment of
inhomogeneous stellar atmospheres (Stuik, Bruls \& Rutten
1997). However, standard models are still in trouble because they
cannot explain the solar $A({\rm Li})$ of $1.16\pm0.10$, two orders of
magnitude below stars of similar mass and metallicity in the Pleiades.
Non-standard models predict extra Li depletion during MS evolution and
might explain the solar and Hyades data, but predict little extra PMS
depletion and cannot explain the Pleiades Li dispersion (Chaboyer,
Demarque \& Pinsonneault 1995 [C95]). To counter this, and address a
correlation between {\em reduced} Li depletion and rapid rotation in
the Pleiades (S93), Mart\'{i}n \& Claret (1996 [MC96]) suggest
structural changes in {\em fast} rotators lead to inhibited PMS Li
depletion. Alternatively, Ventura et al. (1998) [V98] suggest that PMS
depletion using turbulent convective models can account even for the
solar Li depletion and that depletion inhibition in fast rotating stars
could be accomplished with dynamo generated magnetic fields at the CZ
base.

To isolate the importance of non-standard mixing, we have obtained Li
abundances for G and K stars in Blanco 1, a cluster with a similar age
(50-100\,Myr) to the Pleiades, but a spectroscopically
determined metallicity that has been reported to be greater than the
Hyades ([Fe/H]$=+0.23$ -- Edvardsson et al. 1995 [E95]), although we
discuss this point at some length in this paper. Current standard
models predict more Li depletion than in the Hyades, but if the
suggestions of MC96 or V98 are correct, there may be a rotation-rate
dependent scatter above this low level. Some observations of Li from
low resolution spectra have been presented previously by Panagi \&
O'Dell (1997). Their data were seriously incomplete and of relatively
low signal to noise (S/N), but did seem to suggest that there may be at least
some G and K stars in Blanco 1 with as much Li as the Pleiades and
considerably more than the Hyades.

\section{Observations and Results}

G and K stars in Blanco 1 were selected from photometric candidates in
Panagi \& O'Dell (1997). We emphasize that these were chosen on the
basis of their position in the colour-magnitude diagram (CMD); not
their chromospheric activity, which might have biased the sample towards
fast rotators. High resolution spectroscopy was obtained using the
3.9-m Anglo Australian Telescope with the University College London
Echelle Spectrograph on the nights of 20-21 September 1997. A 79
lines/mm echelle was used with a 1024x1024 Tektronix CCD as the
detector. Partial order coverage from about 5400\AA\ to 8400\AA\ was
obtained. A 1.2 arcsec slit projected to three 24$\mu$m pixels, giving
an effective resolution of 0.15\AA. Exposure times were 800\,s to
3000\,s, giving an optimally extracted S/N of $\sim30$ per pixel around
the \lii\ 6707.8\AA\ feature. Data reduction and analysis were
performed in a similar way to that described in James \& Jeffries (1997
[JJ97]) and only brief details are provided here.

Heliocentric radial velocities ($RV$) and projected equatorial
velocities (\vsini) were obtained by cross-correlation with IAU $RV$
standards and slowly rotating stars of similar spectral type (see
JJ97). Errors in $RV$ were $\pm$(0.2-2)\kms, depending on \vsini.
\vsini\ was determined to $\sim10$ percent down to a lower limit of
5\kms, set by the instrumental resolution and S/N. In this paper we
concentrate on the 17 stars (out of 33 observed) which remain, after
iteratively rejecting stars that are more than $1.6\sigma$ away from
the weighted mean, after adding a typical true cluster dispersion of
0.7\kms\ in quadrature to their $RV$ errors (Rosvick, Mermilliod \&
Mayor 1992). The final cluster weighted mean $RV$ we adopt is
$5.0\pm0.2$\kms. The extra external zero point error, based on the $RV$
standards HR\,1829, HR\,6349 and HR\,6970, is probably less than about
0.5\kms. Our mean $RV$ is similar to the $3.9\pm0.7$\kms\ deduced by
E95, although see section~3 for a discussion of their adopted
membership criteria. Data for the candidate $RV$ members are presented
in Table~1. Data for the other 16 $RV$ non-members are given in
Table~2.

The equivalent widths (EWs) of the \lii\ 6707.8\AA\ (unresolved)
doublet were calculated by subtracting the spectra of old, Li-depleted
standard stars (see JJ97 for details) of similar spectral type
(broadened if necessary) and integrating the residual Li feature over
$\pm$(\vsini\ + the FWHM of an unresolved line).  This has the effect of
removing (at least to first order) any contribution from a nearby \fei\
line at 6707.44\AA.  LTE Li abundances were calculated by interpolating
the curves of growth given by S93.  NLTE corrections were applied to
these using the code of Carlsson et al. (1994). Effective temperatures
were derived from photometry using the metallicity dependent Saxner \&
Hammarb\"{a}ck (1985 [SH85]) calibration for $(B-V)_{0}<0.64$. For cooler
stars we apply the B\"{o}hm-Vitense (1981) solar metallicity
calibration, modified for metallicity in the same way as the SH85
relation. This is an approximation, but agrees
reasonably well with theoretical results for cool stellar
atmospheres (Kurucz 1993). $B-V$ values were taken from the photographic
values of de Epstein \& Epstein (1985) or from the photoelectric
measurements of Westerlund et al. (1988) if available. For Blanco~1, 
a reddening of
$E(B-V)=0.02$ and [Fe/H]$=+0.14$ were assumed (see section 3.1).  
Errors in these assumed
values have some affect on the derived Li abundances. However, changing
the metallicity, reddening or effective temperature scale essentially
shift the Li abundances {\em along} the direction of the trend in the
data and models (see Fig. 1) and are therefore not crucial in judging
the overall depletion of Li in Blanco 1 with respect to models or other
clusters.  The assumed metallicity is of course vital in discussing the
significance of this depletion as we will shortly see.

Because most stars have only photographic photometry we checked for
systematic errors by plotting a CMD of the probable cluster
members. There are no systematic deviations from a ZAMS locus,
indicating that the colour calibration is adequate. We also estimated
EWs for many weak neutral metal lines in the $RV$ orders of the
probable cluster members.  The scatter around a smooth EW-$T_{\rm eff}$
relation is consistent with the EW errors if the photographic $B-V$
indices have errors of $\pm 0.03$, in agreement with the estimates of
de Epstein \& Epstein (1985).  Errors in the photoelectric $B-V$
indices are assumed to be $\pm0.01$. Table~1 contains the magnitudes,
colours, measured and derived quantities for our sample of $RV$ cluster
members. Stars are identified with the ZS numbers given them by de
Epstein \& Epstein. Table~2 contains the same data for the $RV$
non-members, although note that if their reddening or metallicity
values are not consistent with Blanco 1, then the Li abundances will be
in error.  There is some evidence from Li abundances and also magnetic
activity (Panagi \& O'Dell 1997), that several of these stars probably
{\em are} cluster members in binary systems -- for instance ZS\,83,
ZS\,154 and ZS\,165. As we have only one $RV$ measurement for these
objects we cannot be sure, and we defer a full discussion of these
objects to a later paper.  Fig.~1 shows the Li abundances of the $RV$
members in Blanco 1 compared with abundances in the Pleiades and
Hyades.  \lii\ 6707.8\AA\ EWs and $B-V$ values for these clusters were
taken from S93, T93, Boesgaard \& Budge (1988) and Soderblom et
al. (1995) and converted to abundance in the same way as the Blanco 1
data, assuming the metallicities quoted in section 1.

The G and K stars of Blanco 1 have Li abundances at a similar level to
the Pleiades and show considerably less depletion than the Hyades.
There is also a scatter in the \lii\ 6707.8\AA\ EW, $B-V$ relationship
among the K stars of Blanco 1, which is similar to that in the
Pleiades. This large dispersion could not be caused by photometry
errors alone. This is illustrated in Fig.~2 where we show normalized
spectra for ZS\,44 and ZS\,243 and the difference between them.  All
the spectral features except \lii\ have cancelled out. We believe that
whatever is responsible for the apparent dispersion in the Pleiades Li
abundances is also at work in Blanco 1.

For the 14 stars in Table 1 with ($0.72<B-V$)$_{0}<1.04$, there are 8
with $\vsini \leq 6$\kms, 3 with $6 < \vsini \leq 10$\kms\ and 3 with
$\vsini > 10$\kms. Queloz et al. (1998) present resolved \vsini\
measurements for several hundred cool Pleiades stars.  There are $\sim45$
percent of Pleiads in a similar ($B-V$)$_{0}$ range with $\vsini \leq
6$\kms, 30 percent with $6 < \vsini \leq 10$\kms\ and the rest have
$\vsini > 10$\kms\ with a small tail extending to $\sim 100$\kms.  In
the Hyades the {\em true} equatorial velocities for stars in this
colour range lie in a narrow band of 3-5\kms\ (Radick et al. 1987).
The Blanco 1 late G and K stars are reasonably consistent with the
Pleiades \vsini\ distribution and there are enough fast rotators for us
to claim that spin down to the Hyades distribution is required.  We
have \vsini\ in too few early G stars to make a sensible comparison
with the Pleiades, but with $\vsini=11$\kms, ZS\,58 rotates faster than
any similar star in the Hyades.

% Authors may indicate to the editorial staff where they would like 
% figures and tables to be placed in the manuscript.  This is done with
% either the \placefigure{KEY} or \placetable{KEY} commands.  These
% commands require \label{KEY} commands to be placed appropriately with
% corresponding table and figure captions.  When the manuscript is
% printed a short note is printed on the page where the figure or table
% is to go.  These commands are ignored in the aaspp4 and aas2pp4 styles.

%\placetable{tbl-3}
%\placefigure{fig1}

% In this section, we see the use of the \subsection command to set off
% an independent subsection.  We only have one here; usually there would
% be several.

% We show the use of several of the displayed math environments described
% in the User Guide, and you get a healthy dose of mathematical typesetting
% examples.  Also, observe the use of the LaTeX \label command after the
% \subsection to give a symbolic KEY to the subsection for cross-referencing
% in a \ref command.  LaTeX automatically numbers the sections, equations,
% tables, etc., as it goes, so in general you don't know what number something
% is going to have.  We'll refer to the "hairymath" section a little later.

\section{Discussion}

The interpretation of the Blanco 1 Li data depends upon the assumed age
and crucially, the assumed metallicity.  The age of Blanco 1 is
uncertain because of a sparse upper main sequence. Age estimates range
from 50\,Myr (de Epstein \& Epstein 1985), based upon fitting the
positions of high mass members in the Hertzsprung-Russell diagram, to
$90\pm25$\,Myr (Panagi \& O'Dell 1997), based upon the chromospheric
activity exhibited by its K dwarfs. In either case, the lowest mass
stars considered here are old enough to have undergone their entire PMS
Li depletion ({\em e.g.} D'Antona \& Mazzitelli 1994).  For that reason
we compare Blanco~1 with the slightly older, but well studied Pleiades
cluster, although there would be no change in any of our arguments
were we to compare it with the younger Alpha Persei cluster (age 
52\,Myr -- Meynet et al. 1993, [Fe/H]$=-0.057\pm0.051$ -- BF90), which
has an almost identical Li depletion pattern to the Pleiades
(Balachandran et al. 1988, 1996).
If we were to assume that the [Fe/H] values for Blanco 1, the
Hyades and Pleiades are $+0.23$ (or $+0.14$ -- see section 3.1), 
$+0.127$ and $-0.034$ and that all
other elements scale with solar abundances, then our results would have
profound implications for PMS Li depletion/preservation and
non-standard MS mixing (see below).  If for some
reason though, the metallicity of Blanco 1 has been
over-estimated and is closer to the solar value, then it would
be no surprise to see a Li depletion pattern similar to the Pleiades.

\subsection{The cluster metal abundances}

\label{metal}

The conclusions of this paper are crucially dependent on the relative
abundances of Blanco~1, the Pleiades and Hyades.  The Pleiades and
Hyades [Fe/H] values we adopt are from high resolution spectroscopy
performed by BF90 and consistent results were published by Cayrel,
Cayrel de Strobel \& Campbell (1985) and Boesgaard (1989). About 15
\fei\ lines were used, $T_{\rm eff}$ was derived for the Pleiades from
$uvby$ and $B-V$ photometry and based upon the SH85
calibrations.  For the Hyades, the Cayrel et al. (1985) $T_{\rm
eff}, V-K$ calibration was adopted. A $\log g$ of 4.5 and
microturbulence, $\mu\sim 1.1-1.4$\kms, were assumed.  14 well known F
dwarf members of the Hyades and 12 F dwarfs in the Pleiades yielded
[Fe/H] values of $+0.127\pm0.022$ and $-0.034\pm0.024$ respectively.
The spectroscopic abundance analysis of Blanco~1 was performed by E95
with high resolution, high s/n spectra of only four F stars in the
cluster.  They found [Fe/H]$=+0.23\pm 0.01$ (p.e.) and an average
metallicity for iron peak elements [M/H]$=+0.18\pm0.04$. Both neutral
and ionized Fe lines were used to check $T_{\rm eff}$ values derived
from $uvby$ photometry and the calibration of Edvardsson et al. (1993).
A $\log g$ of 4.5 and microturbulence, $\mu=$2.0-2.5\kms, were assumed.

The two primary areas of concern that must be addressed before we can
safely accept the abundances of E95 are (i) the differences in the assumed
temperature scales and $\mu$ between their work and the investigations
of BF90 for the Hyades and Pleiades and (ii) whether the 4 F dwarfs
chosen by E95 are actually members of the Blanco 1 cluster.

If the $\mu$ adopted for the Hyades and Pleiades were used for Blanco 1
then even higher metal abundances would be deduced. However, $T_{\rm
eff}$ would also have to be reduced by 600\,K to keep abundances
derived from lines with different excitation energies constant
(E95). This would be very inconsistent with the photometric $T_{\rm
eff}$ values and yield very different abundances for neutral and
ionized lines.  As BF90 measured line EWs which were almost independent
of $\mu$, it seems more likely that their $\mu$ could be increased to
match that assumed for Blanco 1 with little difference in the relative
abundances. An interesting comparison of the temperature calibrations
used in these papers is presented by Balachandran (1995). The $T_{\rm
eff}$ values used for the Hyades by BF90, based on the Cayrel et
al. (1985) calibration, are similar to the E95 temperatures for the
same $b-y$ at $\sim 6800$\,K, but about 50\,K cooler at
$\sim6300$\,K. The Cayrel et al. temperatures match the SH85 
temperatures used by BF90 for the Pleiades at
$\sim 6300\,K$ but are 150\,K hotter at 6800\,K.  Because the Blanco 1
F stars are hotter on average than the F stars used to obtain
metallicities in the Pleiades and Hyades, we estimate that if all these
stars were placed on a uniform temperature scale 
(the SH85 scale for instance), then the Blanco 1 metallicities
should be reduced by about 0.09 dex, the Hyades metallicities should be
reduced by only $\sim0.03$ dex and the Pleiades would remain the same (see
E95 or Balachandran 1995 for example calculations). Thus even with
these modifications it would still seem likely that Blanco 1 was more
metal rich than the Pleiades and probably than the Hyades as well.  We
also note that a reduction of 150\,K in the Blanco 1 temperatures would
increase the [Fe/H] derived from ionized lines and bring them into
better agreement with the results from neutral lines. The results in
Tables~1, Table 2 and Fig.1 therefore assume that [Fe/H]$=+0.14$ for
Blanco 1. 

A second possible problem is that the four F stars chosen by E95 to
represent the Blanco 1 cluster may have questionable cluster membership
credentials. A number of flaws in the reasoning of E95 in selecting
these stars can be found and to aid this discussion we refer the reader
to Tables 2 and 6 in E95. First, membership is assigned on the basis of
distance moduli, $RV$ and chemical abundance. The latter is circular
reasoning -- two stars in E95's sample are excluded (ZS\,28 and ZS\,60)
because they have near solar abundances, but are otherwise excellent
candidate members. For ZS\,60, which has an [Fe/H] of $+0.02$, E95
suggest that rotational broadening may have caused under-estimates of
line EWs, but there is no proper justification given for excluding ZS\,28,
which has an [Fe/H] of $+0.10$. Secondly, unweighted averages are taken
of $RV$ measurements from different sources with different
errors. These are then used to try and find a common cluster $RV$, in
much the same way as we have done in this paper but with fewer
stars. Two of the four stars finally selected as members are $RV$
variables (ZS\,70 and ZS\,125) and the average $RV$ for these objects
has been found from only three measurements. It could be argued that of
the stars with stable and consistent $RV$s, the most likely members in
E95 are ZS\,8, ZS\,28 and ZS\,60, with an average $RV$ of $4.8$\kms\
that is similar to the mean $RV$ derived in this paper, although ZS\,8
was ruled out on the grounds of its distance modulus. The two single
stars selected as members 
by E95 are ZS\,47 and ZS\,63 which have {\em weighted}
mean $RV$s of 0.4\kms\ and 3.8\kms\ respectively. Neither of these
would have been selected as single cluster members according to the
criterion adopted by us in the last section (although considering
possible external errors, ZS\,63 is a reasonable candidate). If in fact
ZS\,28 and ZS\,60 are true cluster members and ZS\,47, ZS\,63, ZS\,70
and ZS\,125 were not, then it might be possible that Blanco 1 has a
nearly solar metallicity.

While the above problems should rightly be pointed out and lead us to
be cautious in our conclusions, we believe there are two final pieces
of evidence that do support a high metallicity value for Blanco 1.
Firstly, the eight Population I F stars studied by E95 have very small
{\em internal} [Fe/H] errors of only $\sim 0.01-0.04$ dex, based upon
between 11 and 51 individual lines. Six of these stars, including the
four classified as members by E95 plus ZS\,57 with a varying radial
velocity and ZS\,8 with a discrepant distance modulus, have [Fe/H]
within $2\sigma$ of $+0.23$. Recalling that $RV$ alone is not a
foolproof way of selecting all cluster members, because of the
possibility of SB1 systems (which would display the same [Fe/H] --
E95), this clustering of the [Fe/H] values within a range much smaller
than the general spread of abundances seen in field F stars is
an unlikely coincidence unless the stars are genuinely
physically associated. Secondly, we have checked for proper motion
information in the Hipparcos catalogue (Perryman et al. 1997) in the
region of sky around the Blanco 1 cluster. A number of bright B, A and
F stars have proper motions and there is a clear cluster present in the
$\mu_{\alpha}\cos \delta$ vs $\mu_{\delta}$ plot (see Fig.~\ref{fig3}).
Only two of the Population I F stars with abundances in E95 have proper
motion data and they are labelled in Fig.~\ref{fig3}. ZS\,47 is
classified as a member by E95, with [Fe/H]$=+0.25\pm 0.04$. ZS\,28 is
unconvincingly classified as a non-member by E95 because of its lower
[Fe/H] of $+0.10\pm 0.02$ (see above). However, the evidence from the
proper motions in Fig.~3 {\em supports} these classifications. ZS\,47
is an excellent cluster candidate, whereas ZS\,28 is about 10\,mas/yr
(and many $\sigma$) away from the apparent cluster proper motion, where
1\,mas/yr corresponds to $\sim1.2$\kms\ at the distance of Blanco 1.

\subsection{The possible consequences of the Blanco 1 Li abundances}

If we take the view that Blanco 1 has an [Fe/H]$\sim +0.14$ (assuming
the SH85 temperature scale) compared to
a Hyades [Fe/H] of $\sim+0.10$ and a Pleiades [Fe/H] of -0.03, then the
data in Fig.~1 present severe problems for standard evolutionary models.
According to all these models ({\em e.g.}
D'Antona \& Mazzitelli 1994, S94a,b, C95, P97, V98), 
the G and K stars of Blanco 1 should have
suffered {\em more} PMS Li depletion than either the Pleiades or
Hyades because of their higher metallicity and thicker CZs.
We could then draw the following conclusions:

(1) A mechanism for inhibiting PMS Li depletion is required.  Standard
Li depletion isochrones for [Fe/H]$=+0.15$ are given by P97 (see
Fig.1). PMS Li depletion is 0.3 to 1.0 dex greater than seen in Blanco
1 at 5000-5300\,K, with perhaps 
less difference at higher temperatures. If [Fe/H] is
larger than $+0.14$ for Blanco~1, or the full spectrum turbulence
convective model is adopted (V98) or one assumes the Hyades define an
empirical standard Li depletion isochrone for [Fe/H]$=+0.127$ (S94a,b),
then these discrepancies are even larger. Therefore, inhibited PMS Li
depletion is needed to explain the Blanco 1 results. We believe the
mechanism of MC96, involving structural changes caused by rapid
rotation is not favoured. The slowest rotators in clusters should be
almost unaffected and would have approximately the PMS Li depletion
predicted by standard models. While this could be the case for the
Pleiades it is not for Blanco 1. Suppression of convection by dynamo
generated magnetic fields at the CZ base holds more promise (V98), as
even the slowest rotators in young clusters like the Pleiades and
Blanco 1 are magnetically active.

(2) Extra mixing and Li depletion is required whilst stars are on the
MS, if one is to explain how Li in Blanco 1 can evolve to resemble the
depletion pattern seen in older clusters with similar or lower
metallicity. The required depletion increases as the mass decreases and
must be $>2$ orders of magnitude in $\sim 600$\,Myr at
5200\,K. Standard convective mixing should be ineffective on the MS for
$T_{\rm eff}> 4700$\,K, but non-standard models incorporating diffusion
and rotational mixing {\em can} provide extra depletion (Charbonnel et
al. 1994, C95).  If non-standard mixing is effective in young open
clusters, then it would seem likely that the same physics will operate
in halo stars, with consequences for the determination of Li$_{\rm p}$
from Population\,II Li abundances.  For instance, Chaboyer \& Demarque
(1994) show that changing from standard to non-standard models (with
the same input physics as C95) requires an increase in Li$_{\rm p}$
from $A({\rm Li})=2.2$ to $\sim3.0$ in order to account for present day
Population\,II Li abundances.

(3) Metallicity is a less important parameter than age in determining
the level of Li in young G and K stars, contrary to the predictions of
standard models where the level of Li depletion should be ``frozen in''
at a metallicity dependent ZAMS level. This conclusion is now supported
by data from a number of open clusters (see P97 for a review and
Jeffries, James \& Thurston 1998) with ages between 30\,Myr and
700\,Myr. Although much of the cluster metallicity information is
uncertain or determined from photometry, the trend is clearly of
increasing depletion with age and there are no exceptions to this.
Metallicity may be apparent as a second order effect as clusters become
older. There is evidence of metallicity dependent Li differences
between the Hyades, Praesepe and NGC 6633 at 600-700\,Myr (Jeffries
1997).  Such differences would be expected from non-standard mixing,
which depends on the distance between CZ bases and the depth at which
effective Li burning commences (C95). The lack of metallicity
dependence in younger stars means that compositional inhomogeneities
could not explain star to star scatter of Li abundances in stars of the
same age and that Li abundance remains a reasonable age estimator for
young ($<700$\,Myr), cool Population\,I stars with unknown metallicity.

There are however two caveats that must be addressed before these
conclusions can be confidently accepted. The first is the lingering
doubts about the metal content of Blanco 1, which have been partially
assuaged above, but really necessitate an extensive study of F stars in
the cluster to remove the remaining uncertainties. Until that is done,
one could still take the point of view (ignoring the results of E95 in
the process) that the standard models of Li mixing and depletion are
largely correct and that our results for Blanco 1 actually demonstrate
that the cluster has a near solar metallicity!

A second possibility which could also be addressed with new, high
resolution F star observations is that S94a show that knowledge of
[Fe/H] alone, does not necessarily lead to accurate predictions of PMS
Li depletion in standard models.  Si, Ne and O are also major sources
of opacity at depths important for Li depletion. S94b cite a private
communication from J. R. King (1993) that [O/H] for the Hyades is
$+0.265\pm0.05$ and show that this leads to 0.4 dex more PMS Li
depletion at 5200\,K compared with a model for which [O/H]$=$[Fe/H].
If, unlike Fe, the Si, Ne and O abundances in Blanco 1 were similar or
less than those in the Pleiades, then it may yet be possible to explain
the observed Li depletion patterns in terms of standard models. The
only one of these abundances we have is [Si/H]$+0.13\pm0.05$ for Blanco
1 (which should be reduced to $+0.07$ if we use the SH85 $T_{\rm eff}$
calibration -- E95), which is smaller than [Fe/H], but not as small as
[Fe/H] in the Pleiades.

\section{Summary}

Examining the assumptions behind the metal abundance determinations in
Blanco~1, the Hyades and Pleiades, we conclude that Blanco 1 is
probably metal-rich with respect to the Pleiades and more similar to
the Hyades.  If this can be confirmed, standard (convective mixing
only) evolutionary models predict that the Blanco 1 G~and~K stars
should have suffered even more PMS Li depletion than the Hyades. Our
observations show that the Li abundances and rotation rates in Blanco 1
are similar to the Pleiades and much higher than the Hyades. To explain
this would require mechanisms that inhibit PMS Li depletion and then
provide non-standard mixing and Li depletion whilst stars are on the
MS. The extra mixing must cause Li depletion amounting to at least 2
orders of magnitude at 5200\,K in 600\,Myr, but less at higher
temperatures.  The rotational mixing models proposed by a number of
authors seem capable of providing this.  These findings could also be
good news for the use of Li as an age indicator in field stars. For
stars younger than the Hyades, modest metallicity differences do not
seem to result in large differences in Li depletion.  Finally, if
non-standard mixing is/was at work in Population\,II stars then the
primordial Li abundance could be significantly higher than their
presently measured Li abundances, with consequences for big bang
nucleosynthesis and the baryonic density.

We add a note of caution to these conclusions that there are still some
question marks over the metallicity determination in Blanco 1 and it is
also possible that non-solar abundance ratios of Fe, O, Si and Ne 
may yet make the Li depletion pattern of Blanco 1 more consistent with
standard evolution models. There is an urgent need for  detailed high
resolution observations of a number of F/G stars in Blanco 1 in order to
address these issues.

% The \notetoeditor{TEXT} command allows the author to communicate some
% information to the copy editor.  This information will appear as a 
% footnote on the printed copy for the aasms4 style file.  Nothing will 
% appear on the printed copy if the aaspp4 or aas2pp4 style file is used.

% In these sections, we see some additional math-related markup, and we
% have references to one of the tables (occurs later in the document)
% and the "hairymath" section immediately preceding this one.
%
% In the second paragraph, note the use of in-text math ($stuff$) including
% a couple of the miscellaneous symbol commands defined in the AASTeX macro
% package.
%
% This is the last section of the paper, so there is an \acknowledgments
% section at the end of the main body.

\acknowledgments

We thank the Director and staff of the Anglo-Australian Observatory for
their assistance in collecting the data. Computational work was done at
the Keele and St Andrews Starlink nodes, funded by the UK Particle
Physics and Astronomy Research Council. We thank an anonymous referee
for a careful reading of the initial manuscript and urging a more
critical evaluation of the Blanco 1 metallicity.

\clearpage

\begin{deluxetable}{lrrrrrrr}
\footnotesize
\tablecaption{Observed and derived parameters for $RV$ members of Blanco 1}
\tablewidth{0pt}
\tablehead{
\colhead{Star\tablenotemark{a}} & \colhead{$V$\tablenotemark{b}} &
\colhead{$B-V$\tablenotemark{b}} & \colhead{\vsini} & \colhead{$RV$} &
\colhead{\lii\, 6708\AA} & \colhead{$T_{\rm eff}$\tablenotemark{c}} & \colhead{$A({\rm Li})$}
\nl & & & \colhead{(\kms)} & \colhead{(\kms)} & \colhead{EW
(m\AA)} & \colhead{(K)} &} 
\startdata
ZS\,31 & 12.43 & 0.75 & $69.4\pm5.0$ & $1.7\pm2.0$ & $232\pm13$ &
5571 & $3.08\pm 0.10$  \nl
ZS\,38 & 13.63 & 1.00 & $69.1\pm6.0$ & $5.5\pm2.0$ & $162\pm18$ &
4988 & $2.27\pm0.11$   \nl
ZS\,44 & 13.19 & 0.86 & $5.5\pm1.5 $ & $5.2\pm0.2$ & $269\pm8$&
5303 & $3.03\pm0.08$   \nl
ZS\,45 & 12.76 & 0.84 & $5.6\pm0.6 $ & $5.2\pm0.2$ & $207\pm5$&
5349 & $2.79\pm0.08$   \nl
ZS\,58 & 12.15 & 0.68 & $11.0\pm0.7$ & $4.7\pm0.4$ & $192\pm5$&
5766 & $3.06\pm0.08$   \nl
ZS\,61 & 13.51 & 0.93 & $56.8\pm5.0$ & $3.7\pm2.0$ & $212\pm15$&
5145 & $2.62\pm0.10$   \nl
ZS\,70 & 13.51 & 0.93 & $\leq 5.0$   & $4.9\pm0.2$ & $85\pm 7$&
5145 & $2.01\pm0.09$   \nl
ZS\,75 & 12.79 & 0.94 & $\leq 6.0$   & $4.9\pm0.2$ & $134\pm6$&
5122 & $2.26\pm0.08$   \nl
ZS\,93 & 13.92 & 0.98 & $\leq 5.0$   & $5.2\pm0.2$ & $81\pm11$ &
5033 & $1.88\pm0.11$   \nl
ZS\,102& 12.50 & 0.77 & $7.3\pm0.7$ & $5.0\pm0.2$ & $141\pm7$ &
5519 & $2.65\pm0.05$   \nl
ZS\,112& 13.02 & 0.90 & $\leq 5.0 $  & $4.6\pm0.2$ & $185\pm6$ &
5212 & $2.57\pm0.07$   \nl
ZS\,135& 13.52 & 0.97 & $6.7\pm1.3$  & $5.7\pm0.3$ & $178\pm10$&
5056 & $2.40\pm0.08$   \nl
ZS\,141& 11.96 & 0.68 & $6.4\pm0.4$  & $5.2\pm0.2$ & $147\pm5$&
5766 & $2.86\pm0.08$   \nl
ZS\,147& 13.26 & 0.91 & $\leq 5.0$   & $4.6\pm0.2$ & $160\pm7$&
5189 & $2.44\pm0.08$   \nl
ZS\,158& 13.47 & 1.05 & $\leq 5.0$   & $5.4\pm0.2$ & $116\pm10$&
4875 & $1.92\pm0.09$   \nl
ZS\,182& 11.72 & 0.63 & $7.8\pm0.7$  & $5.0\pm0.2$ & $134\pm8$&
5941 & $2.94\pm0.05$   \nl
ZS\,243& 13.02 & 0.82 & $\leq5.0$    & $5.5\pm0.2$ & $154\pm8$ &
5397 & $2.60\pm0.08$   \nl
\enddata

\tablenotetext{a}{Identifier used by de Epstein \& Epstein (1985)}
\tablenotetext{b}{Photometry from Westerlund et al. (1988) for ZS\,102
and ZS\,182 or de Epstein \& Epstein (1985) otherwise}
\tablenotetext{c}{$T_{\rm eff}$ assumes that $E(B-V)=0.02$ and 
[Fe/H]$=+0.14$ (see text)}
\end{deluxetable}
\clearpage

\begin{deluxetable}{lrrrrrrr}
\footnotesize
\tablecaption{Observed and derived parameters for $RV$ non-members or
possible spectroscopic binary members of Blanco 1}
\tablewidth{0pt}
\tablehead{
\colhead{Star\tablenotemark{a}} & \colhead{$V$\tablenotemark{b}} &
\colhead{$B-V$\tablenotemark{b}} & \colhead{\vsini} & \colhead{$RV$} &
\colhead{\lii\, 6708\AA} & \colhead{$T_{\rm eff}$\tablenotemark{c}} & \colhead{$A({\rm Li})$}
\nl & & & \colhead{(\kms)} & \colhead{(\kms)} & \colhead{EW
(m\AA)} & \colhead{(K)} &} 
\startdata
ZS\,67 & 11.86 & 0.71 & $\leq5.0$    & $28.0\pm0.2$ & $45\pm12$ &
5679 & $2.15\pm0.13$   \nl
ZS\,83 & 12.50 & 0.92 & $6.8\pm1.1$  & $8.9\pm0.3$ & $134\pm7$&
5167 & $2.30\pm0.08$   \nl
ZS\,100 & 12.49 & 0.88& $\leq5.0$ & $10.7\pm0.3$ & $134\pm11$ &
5258 & $2.38\pm 0.09$  \nl
ZS\,113 & 12.92 & 0.86 & $\leq5.0$ & $-8.6\pm0.2$ & $<26$&
5303 & $<1.59$   \nl
ZS\,151 & 13.03 & 0.87 & $\leq5.0$ & $18.4\pm0.6$ & $<22$&
5281 & $<1.49$   \nl
ZS\,153& 12.38 & 0.73 & $\leq5.0$ & $-13.6\pm0.2$ & $23\pm8$ &
5624 & $1.81\pm0.20$   \nl
ZS\,154 & 13.36 & 0.95 & $\leq5.0$ & $7.2\pm0.2$ & $126\pm9$&
5100 & $2.20\pm0.09$   \nl
ZS\,165 & 12.40 & 0.90 & $\leq 7.0$   & $0.8\pm0.2$ & $141\pm 7$&
5212 & $2.37\pm0.08$   \nl
ZS\,175& 13.60 & 1.00 & $\leq5.0$  & $12.3\pm0.2$ & $<30$&
4988 & $<1.35$   \nl
ZS\,180 & 11.77 & 0.62& $\leq5.0$ &  $-1.5\pm0.2$ & $36\pm6$&
5976 & $2.28\pm0.11$   \nl
ZS\,186& 12.22 & 0.75 & $\leq 5.0 $  & $-0.2\pm0.2$ & $27\pm5$ &
5571 & $1.84\pm0.10$   \nl
ZS\,187 & 13.87 & 0.99 & $\leq 5.0$   & $1.6\pm0.2$ & $<18$&
5011 & $<1.14$   \nl
ZS\,195& 11.82 & 0.66 & $\leq 5.0$   & $3.4\pm0.3$ & $64\pm12$&
5827 & $2.44\pm0.11$   \nl
ZS\,228& 13.24 & 0.87 & $\leq5.0$  & $32.7\pm0.3$ & $<30$&
5280 & $<1.63$   \nl
ZS\,234 & 13.04 & 0.91 & $\leq 5.0$   & $-16.3\pm0.3$ & $<20$ &
5190 & $<1.36$   \nl
ZS\,239& 12.50 & 0.89 & $\leq 5.0$   & $9.0\pm0.2$ & $32\pm6$&
5235 & $1.61\pm0.11$   \nl
\enddata

\tablenotetext{a}{Identifier used by de Epstein \& Epstein (1985)}
\tablenotetext{b}{Photometry from Westerlund et al. (1988) for ZS\,67,
ZS\,186 and ZS\,195 or de Epstein \& Epstein (1985) otherwise}
\tablenotetext{c}{$T_{\rm eff}$ assumes that $E(B-V)=0.02$ and 
[Fe/H]$=+0.14$ (see text)}
\end{deluxetable}

% Tabular data can also be aligned within the LaTeX "tabular" environment.  
% Observe that our tabular environment is embedded within a "center" 
% environment, which is in turn inside a "table" environment.  Exercise for 
% the reader:
%
% Why do you think we used the "table*" environment?
%
% We need the table environment for autonumbering and caption generation,
% which is why it is not enough to have a centered tabular.
%
% Within the tabular environment, please note that we use no vertical
% rules, and the only horizontal rule is the \tableline (*not* an \hline)
% which delimits the column headings from the tabular data.  Also note
% that a couple of the column headings require special annotation, i.e.,
% footnotes for tables.  They are marked and tagged with \tablenotemark.
% \tablenotemarks could be placed on individual data entries as well,
% but be careful not to go berserk doing this.

% Camera-ready tables, produced with either the apjpt4 or aj_pt4 style files,
% can be referenced within a table environment using \dummytable.  This acts
% like a place holder and bumps the table counter.   For this particular
% manuscript, tbl-3 refers to the table in file samp2tbl.tex.

% This is the last table for this paper (as well as the first), so we
% should follow it with a \clearpage.  In order to force all the floating
% tables out of their buffers and onto vertical page lists, we must use
% \clearpage rather than \newpage. 

\clearpage

\clearpage

\figcaption[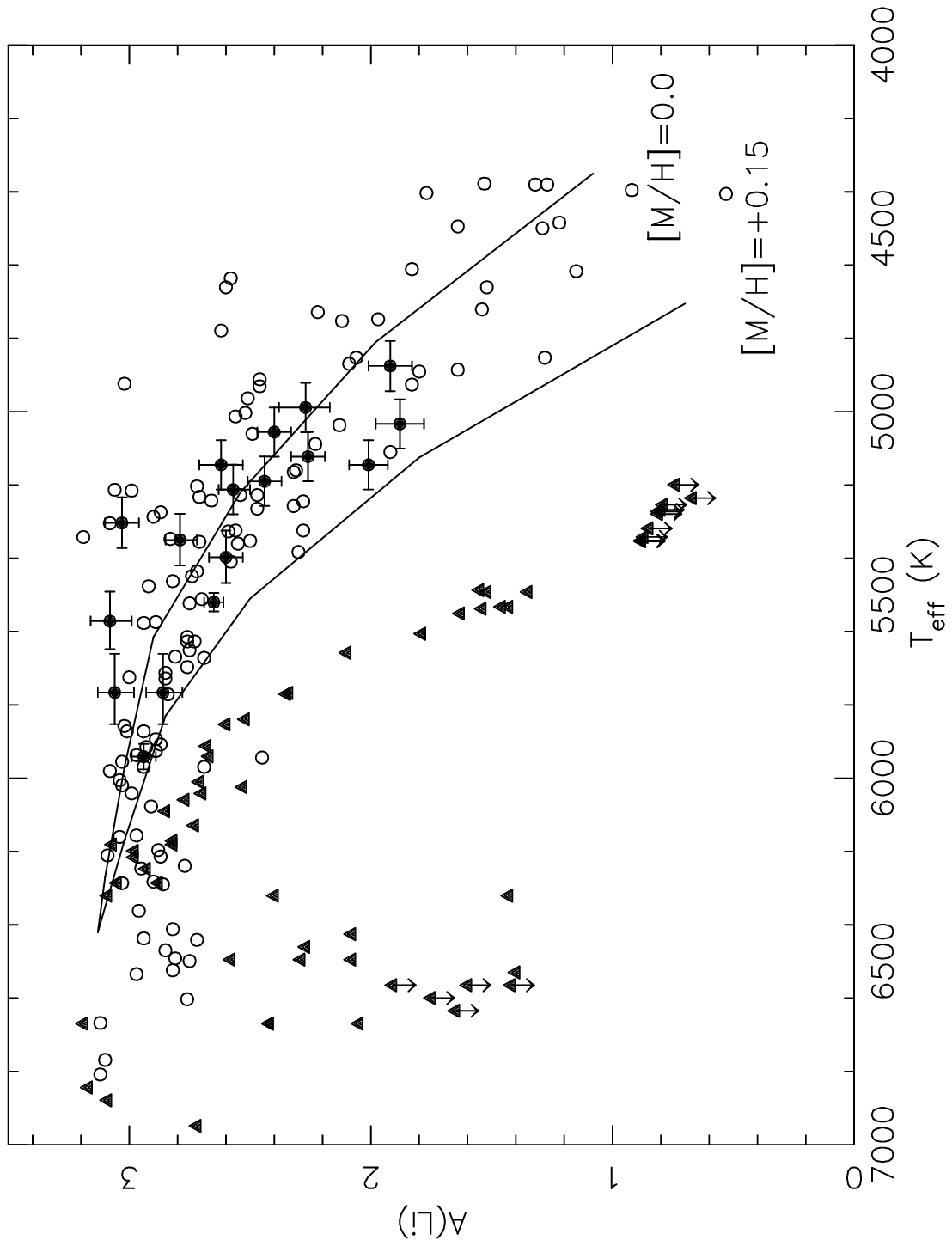]{NLTE lithium abundances, $A({\rm Li})$, for 
Blanco 1 (filled
circles), the Pleiades (open circles) and Hyades (triangles). The solid
lines are standard 100\,Myr Li depletion isochrones from P97, 
assuming the initial $A({\rm Li})$ for Population\,I stars is
3.2. The $A({\rm Li})$ errors for Blanco 1 incorporate both EW and 
$T_{\rm eff}$ errors.
\label{fig1}}

\figcaption[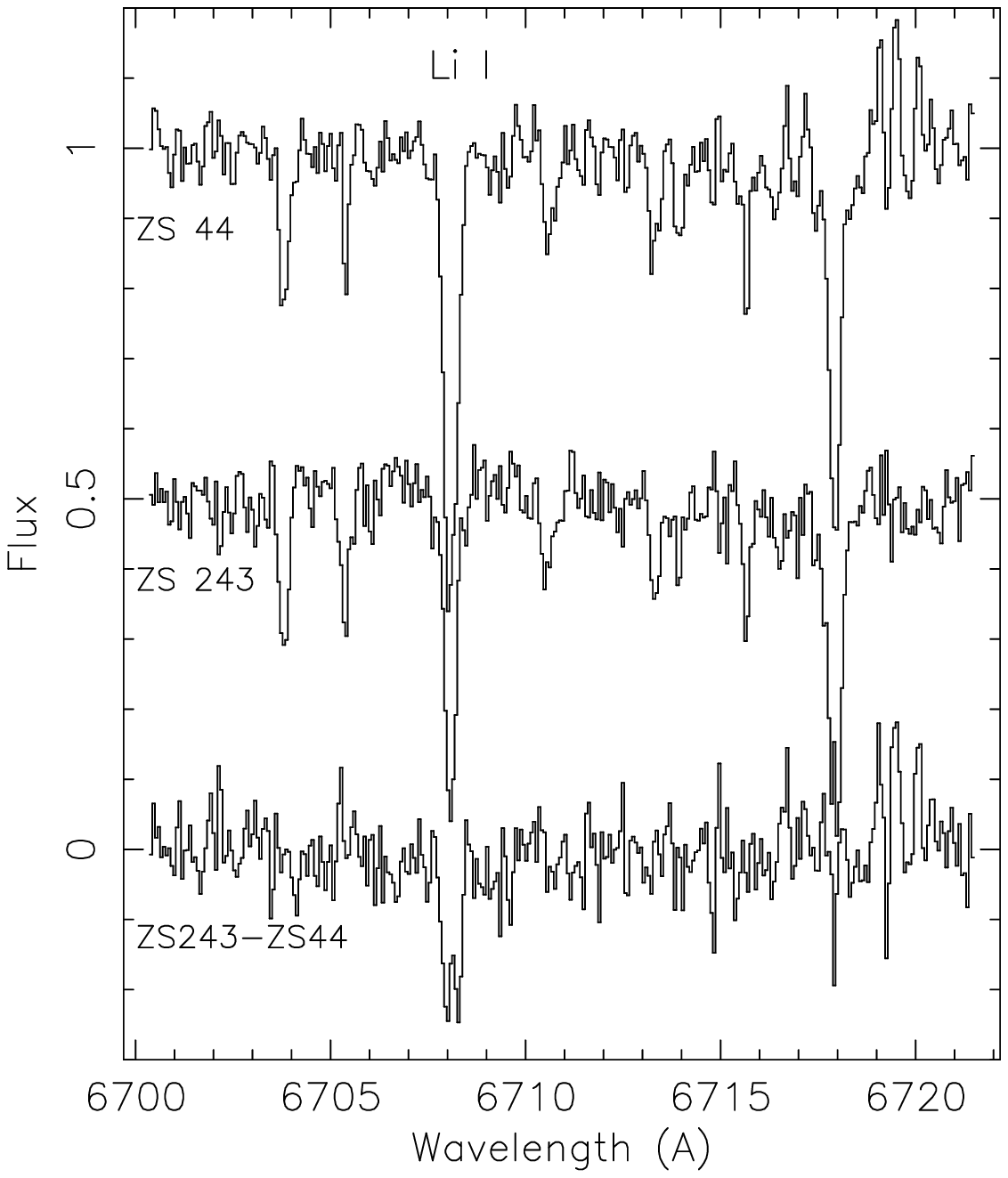]{Normalized spectra around 
the \lii\ 6708\AA\ line for ZS\,44 and
ZS\,243 and the difference between them. The central plot has 
been offset by -0.5.
\label{fig2}}

\figcaption[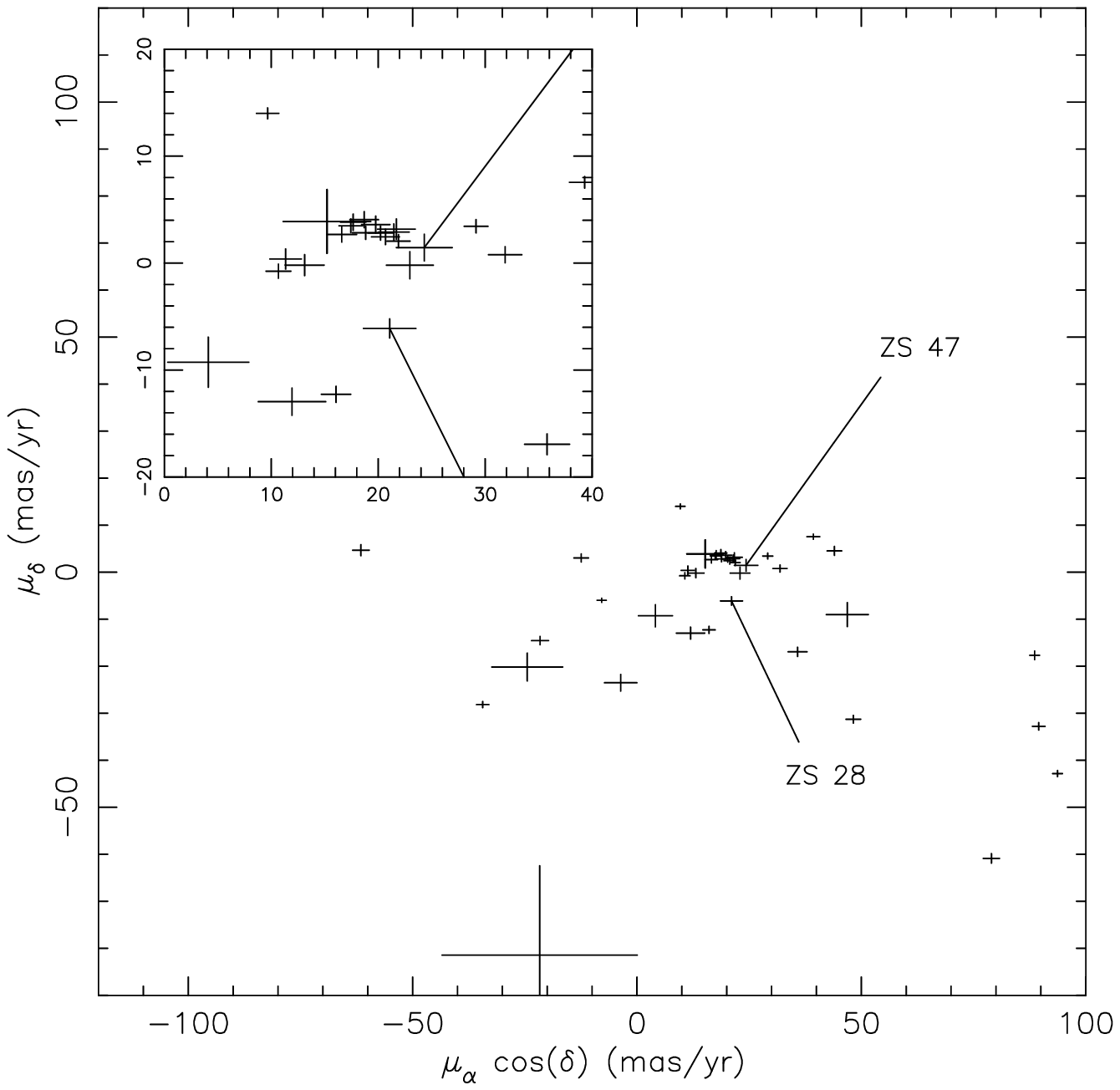]{Proper motions from the Hipparcos catalogue
(Perryman et al. 1997) for stars within 120 arcmin of a nominal Blanco
1 cluster center of 00h02m30s, -30d15m00s (B1950.0). Five stars lie
beyond the borders of the plot. The insert shows a magnification of the
central region of the plot. The two stars marked are discussed in the text.
\label{fig3}}

%\end{document}

% Option 2.  The figure captions are printed on a caption page(s) as in 
% option 1.  The figures available as EPS files are then printed at the
% end of the document, one figure per page, using the \plotone command.
% If you wish to process this option then simply comment out the \end{document}
% just above these five lines. 

\clearpage

\plotone{liplot.eps}

\clearpage

\plotone{lispectra.eps}

\clearpage

\plotone{pmplot.eps}


\begin{thebibliography}{{Cayrel, Cayrel~de Strobel \& Campbell}{1985}}

\bibitem[\protect\citefmt{Balachandran, Lambert \&
  Stauffer}{1988}]{balachandran88}
Balachandran~S., Lambert~D.~L., Stauffer~J.~R., 1988, ApJ, 333, 267

\bibitem[\protect\citefmt{Balachandran, Lambert \&
  Stauffer}{1996}]{balachandran96}
Balachandran~S., Lambert~D.~L., Stauffer~J.~R., 1996, ApJ, 470, 1243

\bibitem[\protect\citefmt{Balachandran}{1995}]{balachandran95}
Balachandran~S., 1995, ApJ, 446, 203

\bibitem[\protect\citefmt{Boesgaard \& Budge}{1988}]{boesgaardbudge88}
Boesgaard~A.~M., Budge~K.~G., 1988, ApJ, 332, 410

\bibitem[\protect\citefmt{Boesgaard \& Friel}{1990}]{boesgaard90}
Boesgaard~A.~M., Friel~E.~D., 1990, ApJ, 351, 467

\bibitem[\protect\citefmt{Boesgaard \& Steigman}{1985}]{boesgaard85}
Boesgaard~A.~M., Steigman~G., 1985, ARA\&A, 23, 319

\bibitem[\protect\citefmt{Boesgaard {\rm et~al.}}{1998}]{boesgaard98}
Boesgaard~A.~M., Deliyannis~C.~P., Stephens~A., King~J.~R., 1998, ApJ, 493, 206

\bibitem[\protect\citefmt{Boesgaard}{1989}]{boesgaard89}
Boesgaard~A.~M., 1989, ApJ, 336, 798

\bibitem[\protect\citefmt{{B\"ohm-Vitense}}{1981}]{bvitense81}
{B\"ohm-Vitense}~E., 1981, ARA\&A, 19, 295

\bibitem[\protect\citefmt{Bonifacio \& Molaro}{1997}]{bonifacio97}
Bonifacio~P., Molaro~P., 1997, MNRAS, 285, 847

\bibitem[\protect\citefmt{Bouvier, Forestini \& Allain}{1997}]{bouvier97}
Bouvier~J., Forestini~M., Allain~S., 1997, A\&A, 326, 1023

\bibitem[\protect\citefmt{Carlsson {\rm et~al.}}{1994}]{carlsson94}
Carlsson~M., Rutten~R., Bruls~J. H. M.~J., Shchukina~N.~G., 1994, A\&A, 288,
  860

\bibitem[\protect\citefmt{Cayrel, Cayrel~de Strobel \&
  Campbell}{1985}]{cayrel85}
Cayrel~R., Cayrel~de Strobel~G., Campbell~B., 1985, A\&A, 146, 249

\bibitem[\protect\citefmt{Chaboyer \& Demarque}{1994}]{chaboyer94}
Chaboyer~B.~C., Demarque~P., 1994, ApJ, 433, 510

\bibitem[\protect\citefmt{Chaboyer, Demarque \&
  Pinsonneault}{1995}]{chaboyer95}
Chaboyer~B., Demarque~P., Pinsonneault~M.~H., 1995, ApJ, 441, 876

\bibitem[\protect\citefmt{Charbonnel {\rm et~al.}}{1994}]{charbonnel94}
Charbonnel~C., Vauclair~S., Maeder~A., Meynet~G., Schaller~G., 1994, A\&A, 283,
  155

\bibitem[\protect\citefmt{de~Epstein \& Epstein}{1985}]{epstein85}
de~Epstein~A. E.~A., Epstein~I., 1985, AJ, 90, 1211

\bibitem[\protect\citefmt{Deliyannis \& Ryan}{1997}]{deliyannis97}
Deliyannis~C.~P., Ryan~S.~G., 1997, ApJ, 480, L43

\bibitem[\protect\citefmt{Deliyannis, Demarque \& Kawaler}{1990}]{deliyannis90}
Deliyannis~C.~P., Demarque~P., Kawaler~S.~D., 1990, ApJS, 73, 21

\bibitem[\protect\citefmt{Edvardsson {\rm et~al.}}{1993}]{edvardsson93}
Edvardsson~B., Andersen~J., Gustafsson~B., Lambert~D.~L., Nissen~P.~E.,
  Tomkin~J., 1993, A\&A, 275, 101

\bibitem[\protect\citefmt{Edvardsson {\rm et~al.}}{1995}]{edvardsson95}
Edvardsson~B., Pettersson~B., Kharrazi~M., Westerlund~B., 1995, A\&A, 293, 75

\bibitem[\protect\citefmt{James \& Jeffries}{1997}]{james97}
James~D.~J., Jeffries~R.~D., 1997, MNRAS, 292, 252

\bibitem[\protect\citefmt{Jeffries, James \& Thurston}{1998}]{jeffriesn251698}
Jeffries~R.~D., James~D.~J., Thurston~M.~R., 1998, MNRAS, in press

\bibitem[\protect\citefmt{Jeffries}{1997}]{jeffries97n6633}
Jeffries~R.~D., 1997, MNRAS, 292, 177

\bibitem[\protect\citefmt{Kurucz}{1993}]{kurucz93}
Kurucz~R.~L., 1993, Technical Report, Synthe spectral synthesis code,
  Smithsonian Astrophysical Observatory, CD-ROM No. 18

\bibitem[\protect\citefmt{Mart\'{i}n \& Claret}{1996}]{martin96}
Mart\'{i}n~E.~L., Claret~A., 1996, A\&A, 306, 408

\bibitem[\protect\citefmt{Meynet, Mermilliod \& Maeder}{1993}]{meynet93}
Meynet~G., Mermilliod~J.~C., Maeder~A., 1993, A\&AS, 98, 477

\bibitem[\protect\citefmt{Panagi \& O'Dell}{1997}]{panagi97}
Panagi~P., O'Dell~M.~A., 1997, A\&AS, 121, 213

\bibitem[\protect\citefmt{Perryman {\rm et~al.}}{1997}]{perryman97}
Perryman~M. A.~C. {\rm et~al.}, 1997, A\&A, 323, L49

\bibitem[\protect\citefmt{Pinsonneault}{1997}]{pinsonneault97}
Pinsonneault~M.~H., 1997, ARA\&A, 35, 557

\bibitem[\protect\citefmt{Queloz {\rm et~al.}}{1998}]{queloz98}
Queloz~D., Allain~S., Mermilliod~J.~C., Bouvier~J., Mayor~M., 1998, A\&A, in
  press

\bibitem[\protect\citefmt{Radick {\rm et~al.}}{1987}]{radick87}
Radick~R.~R., Thompson~D.~T., Lockwood~G.~W., Duncan~D.~K., Baggett~W.~E.,
  1987, ApJ, 321, 459

\bibitem[\protect\citefmt{Richard, Charbonnel \& Dziembowski}{1996}]{richard96}
Richard~O. aand~Vauclair~S., Charbonnel~C., Dziembowski~W.~A., 1996, A\&A, 312,
  1000

\bibitem[\protect\citefmt{Rogers \& Iglesias}{1992}]{rogers92}
Rogers~F.~J., Iglesias~C.~A., 1992, ApJS, 79, 507

\bibitem[\protect\citefmt{Rosvick, Mermilliod \& Mayor}{1992}]{rosvick92}
Rosvick~J.~M., Mermilliod~J.~C., Mayor~M., 1992, A\&A, 255, 130

\bibitem[\protect\citefmt{Saxner \& {Hammarb\"ack}}{1985}]{saxner85}
Saxner~M., {Hammarb\"ack}~G., 1985, A\&A, 151, 372

\bibitem[\protect\citefmt{Soderblom {\rm et~al.}}{1993}]{soderblom93pleiadesli}
Soderblom~D.~R., Jones~B.~F., Balachandran~S., Stauffer~J.~R., Duncan~D.~K.,
  Fedele~S.~B., Hudon~J.~D., 1993, AJ, 106, 1059

\bibitem[\protect\citefmt{Soderblom {\rm et~al.}}{1995}]{soderblom95hyades}
Soderblom~D.~R., Jones~B.~F., Stauffer~J.~R., Chaboyer~B., 1995, AJ, 110, 729

\bibitem[\protect\citefmt{Soderblom}{1990}]{soderblom90}
Soderblom~D.~R., 1990, AJ, 100, 204

\bibitem[\protect\citefmt{Stuik, Bruls \& Rutten}{1997}]{stuik97}
Stuik~R., Bruls~J. H. M.~J., Rutten~R.~J., 1997, A\&A, 322, 911

\bibitem[\protect\citefmt{Swenson {\rm et~al.}}{1994a}]{swenson94a}
Swenson~F.~J., Faulkner~J., Iglesias~C.~A., Rogers~F.~J., Alexander~D.~R.,
  1994a, ApJ, 422, L79

\bibitem[\protect\citefmt{Swenson {\rm et~al.}}{1994b}]{swenson94b}
Swenson~F.~J., Faulkner~J., Rogers~F.~J., Iglesias~C.~A., 1994b, ApJ, 425, 286

\bibitem[\protect\citefmt{Thorburn {\rm et~al.}}{1993}]{thorburn93}
Thorburn~J.~A., Hobbs~L.~M., Deliyannis~C.~P., Pinsonneault~M.~H., 1993, ApJ,
  415, 150

\bibitem[\protect\citefmt{Ventura {\rm et~al.}}{1998}]{ventura98}
Ventura~P., Zeppieri~A., Mazzitelli~I., D'Antona~F., 1998, A\&A, 331, 1011

\bibitem[\protect\citefmt{Westerlund {\rm et~al.}}{1988}]{westerlund88}
Westerlund~B., Garnier~R., Lundgren~K., Pettersson~B., Breysacher~J., 1988,
  A\&AS, 76, 101

\end{thebibliography}
\end{document}